\documentclass{ws-ijmpa}

\begin{document}

\markboth{Edmond L.~Berger, Jianwei Qiu, and Yili Wang}
{Upsilon Transverse Momentum at Hadron Colliders}

%
\catchline{}{}{}{}{}
%
\title{Upsilon Transverse Momentum at Hadron Colliders\footnote{Argonne report 
ANL-HEP-CP-04-110. Presented at the 2004 Meeting of the APS Division of 
Particles and Fields, University of California, Riverside, CA, Aug 26 - 31, 
2004. To be published in International Journal of Modern Physics A.}}
\author{\footnotesize Edmond L.~Berger\footnote{berger@anl.gov}}
\address{High Energy Physics Division, Argonne National Laboratory\\
Argonne, IL 60439
}
\author{Jianwei~Qiu\footnote{jwq@iastate.edu} \, and Yili~Wang\footnote{yiliwa@iastate.edu, conference speaker}}
\address{Department of Physics and Astronomy, Iowa State University\\
Ames, IA 50010
}
\maketitle
\pub{Received (29 Oct 2004)}{}
\begin{abstract}
We predict the shape of the transverse momentum $p_T$ spectrum of 
$\Upsilon$ production. The distribution at low $p_T$ is dominated  
by the region of small impact parameter $b$ and may be computed  
reliably in perturbation theory.  We resum to all orders in the 
strong coupling $\alpha_s$ the process-independent large logarithmic 
contributions that arise from initial-state gluon showers in the 
small $p_T (\leq M_\Upsilon)$ region.  The cross section 
at large $p_T$ is represented by the ${\cal O} (\alpha_s^3)$ 
lowest-order non-vanishing perturbative contribution. 

\end{abstract}
~\linebreak
Perturbative calculations in quantum chromodynamics (QCD) reproduce 
the transverse momentum distribution of $\Upsilon$ production at 
large $p_T$ at hadron colliders but tend not to address the region 
of low $p_T$ where the bulk of the data is~\cite{cdf}.  The 
distribution at low $p_T$ is strongly influenced by initial-state 
gluon showers.  A fixed-order perturbative treatment in QCD leads 
to $1/p_T^2$-type singular terms enhanced by large higher-order 
logarithmic contributions caused by initial-state gluon radiation.  
These contributions have the form 
$\alpha_s\log^2(M_{\Upsilon}^2/p_T^2)$ for every power of the strong 
coupling $\alpha_s$, and reliable predictions in the
regions of small and moderate $p_T$ require that the logarithmic
contributions be summed to all orders in $\alpha_s$.

This paper is a brief summary of a study reported 
elsewhere~\cite{Berger:2004cc}, along with a new comparison with 
preliminary D\O\ Run~2 data~\cite{d0}.   
We use a two-step factorization procedure to 
represent production of the $\Upsilon$. We assume that a pair of bottom 
quarks $b \bar{b}$ is produced in a hard-scattering short-distance 
process. The pair is produced at a distance scale $\sim 1/(2m_b) \sim 
1/45$~fm, which is much smaller than the physical size of a $\Upsilon$ 
meson. The compact $b\bar{b}$ system is unlikely to become an 
$\Upsilon$ meson at the production point. Instead, the pair must
expand, and the $b$ and $\bar{b}$ will interact with each other 
coherently until they transmute into a physical $\Upsilon$ meson.  
A $b\bar{b}$ pair with invariant mass $Q > 2M_B$ is more likely 
to become a pair of $B$ mesons.  We can write the differential 
cross section in the usual way~\cite{Collins:gx}:  
\begin{equation}
\frac{d\sigma_{AB\rightarrow\Upsilon X}}{dp_T^2 dy}
= \sum_{a,b} \int dx_a\, \phi_{a/A}(x_a)\, dx_b\, \phi_{b/B}(x_b)\, 
\frac{d\hat{\sigma}_{ab\rightarrow\Upsilon X}}{dp_T^2 dy}\, .
\label{hadron:xsec}
\end{equation}
Since the momentum of the heavy quark in the pair's rest frame is 
much less than the mass of the pair, the 
parton-level production cross section might be factored 
further~\cite{Collins:gx}: 
\begin{equation}
\frac{d\hat{\sigma}_{ab\rightarrow\Upsilon X}}{dp_T^2 dy}
\approx \sum_{[b\bar{b}]} 
\int dQ^2 
\left[\frac{d\hat{\sigma}_{ab\rightarrow[b\bar{b}](Q) X'}}
           {dQ^2dp_T^2 dy}
\right]
{\cal F}_{[b\bar{b}]\rightarrow\Upsilon \bar{X}}(Q^2)\, .
\label{parton:xsec}
\end{equation}
The function ${\cal F}_{[b\bar{b}]\rightarrow\Upsilon \bar{X}}(Q^2)$
represents a transition probability for a $b\bar{b}$ pair with quantum numbers 
$[b\bar{b}]$ to transmute into an $\Upsilon$ meson. Different 
models of quarkonium production lead to different choices for ${\cal F}(Q^2)$.  
In a color evaporation (or color-bleaching) model 
(CEM)~\cite{Amundson:1996qr}, 
${\cal F}_{[b\bar{b}]\rightarrow\Upsilon}(Q^2) = C_{\Upsilon}$ if 
$4m_b^2 \leq Q^2 \leq 4M_B^2$.
The non-perturbative constant $C_{\Upsilon}$ sets the overall 
normalization of the cross section.  Its value cannot be predicted.  It  
changes with the specific state of the $\Upsilon$ meson. 

Adopting the Collins, Soper, and 
Sterman (CSS) impact-parameter $b$-space (Fourier conjugate to $p_T$) 
approach~\cite{Collins:1984kg}, we write
\begin{eqnarray}
\frac{d\sigma_{AB\rightarrow [b\bar{b}](Q) X'}^{\rm resum}}
     {dQ^2dp_T^2dy}
&=& 
\int \frac{db}{2\pi}\, J_0(p_T\, b)\, 
b{\cal W}_{AB\rightarrow [b\bar{b}](Q)}(b,Q,x_A,x_B) \, .
\label{resum:QQ}
\end{eqnarray}
The resummed calculation of the transverse momentum distribution 
at low $p_T$ is reliable because the Fourier transform in 
Eq.~(\ref{resum:QQ}) is dominated by the region of small $b$, and is 
not sensitive to the extrapolation to the large $b$~\cite{Berger:2004cc}. 
We keep only the process-independent terms in the Sudakov exponential 
functions. The QCD resummed $p_T$ distribution of $\Upsilon$ production 
can be predicted reliably in the region of small and intermediate $p_T$. 
At larger $p_T$, $p_T \geq M_\Upsilon$, we use the ${\cal O} (\alpha_s^3)$ 
lowest-order non-vanishing perturbative contribution that we 
denote as $\sigma^{\rm pert}$.  
 
Because the $b\bar{b}$ system is not necessarily in a color
neutral state, $\sigma^{\rm pert}$ includes radiation from the heavy
quark system as well as from the incoming partons. 
This final state radiation is not included in $\sigma^{\rm resum}$.
To avoid extrapolation of $\sigma^{\rm pert}$ into region of low
$p_T$, we match the resummed and perturbative components of 
the $p_T$ distribution at a value $p_{T_M}$. As we do
not calculate the order $\alpha_s$ corrections to 
$\sigma^{\rm resum}$, we compensate by increasing $\sigma^{\rm resum}$ by a 
resummation enhancement factor $K_r$ and 
assume that $K_r$ is a constant.

\begin{figure}[ht]
\centerline{\psfig{file=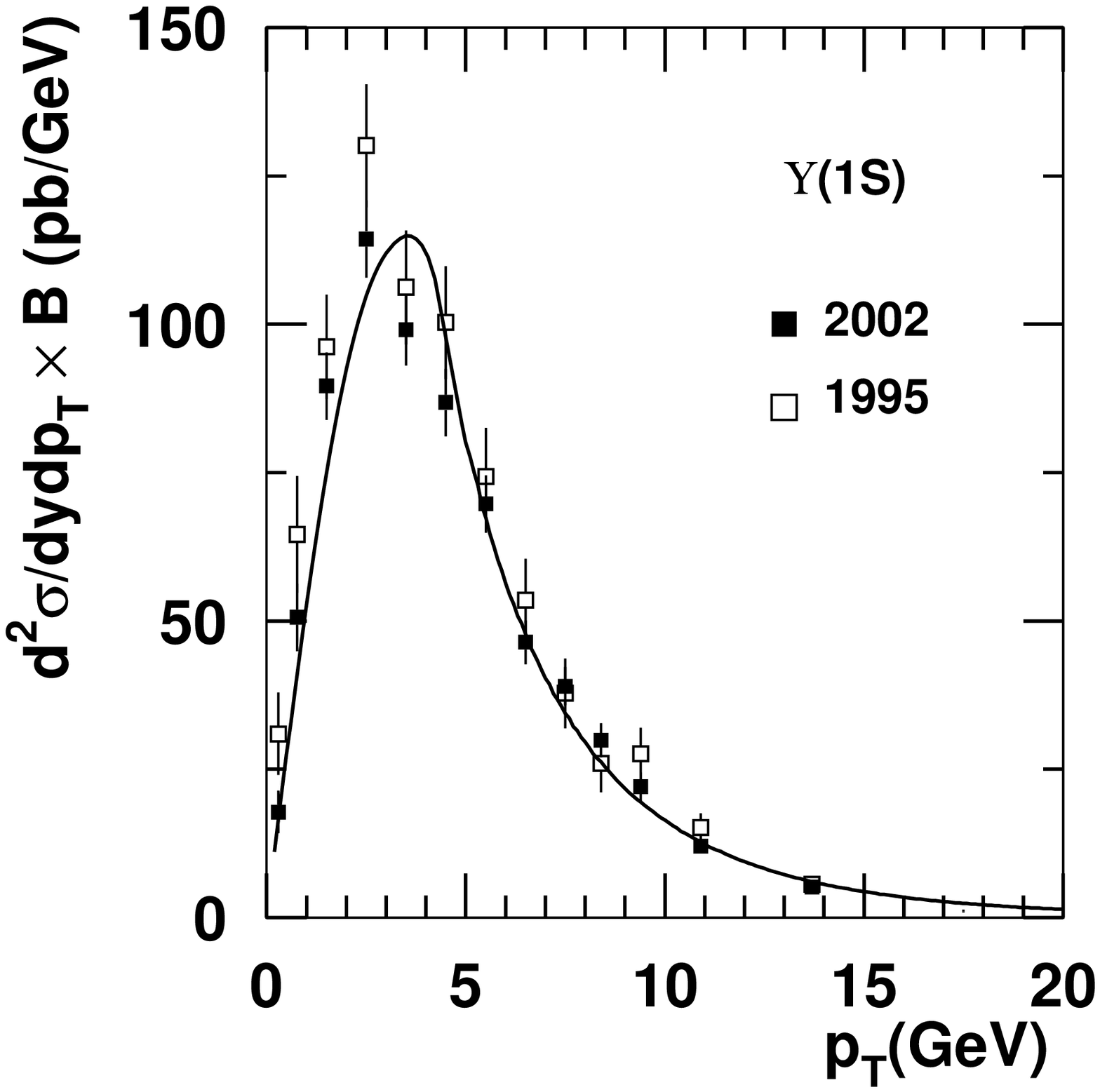, width=3.4cm}
\hspace{0.01cm}      \psfig{file=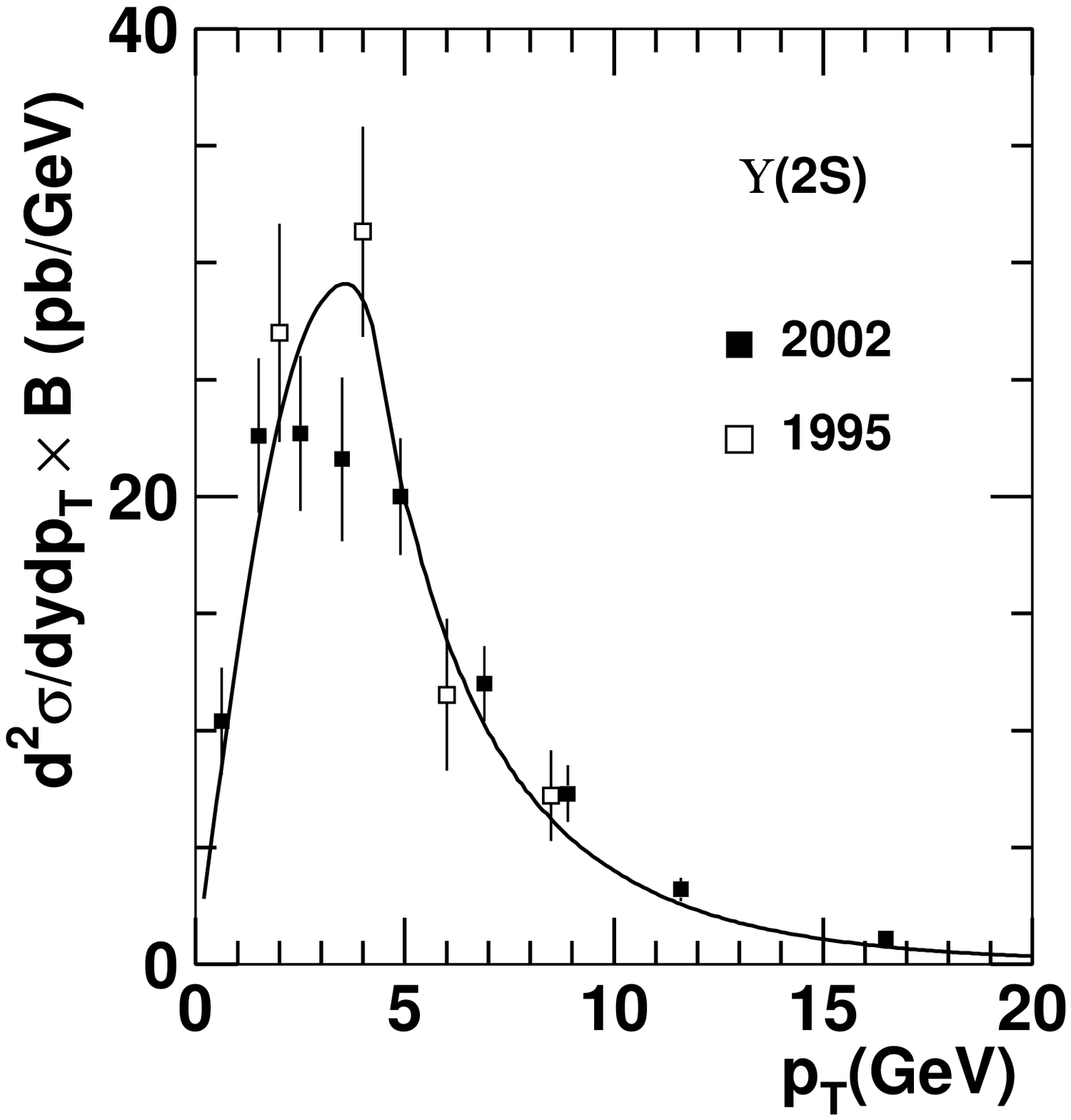, width=3.4cm}
\hspace{0.01cm}      \psfig{file=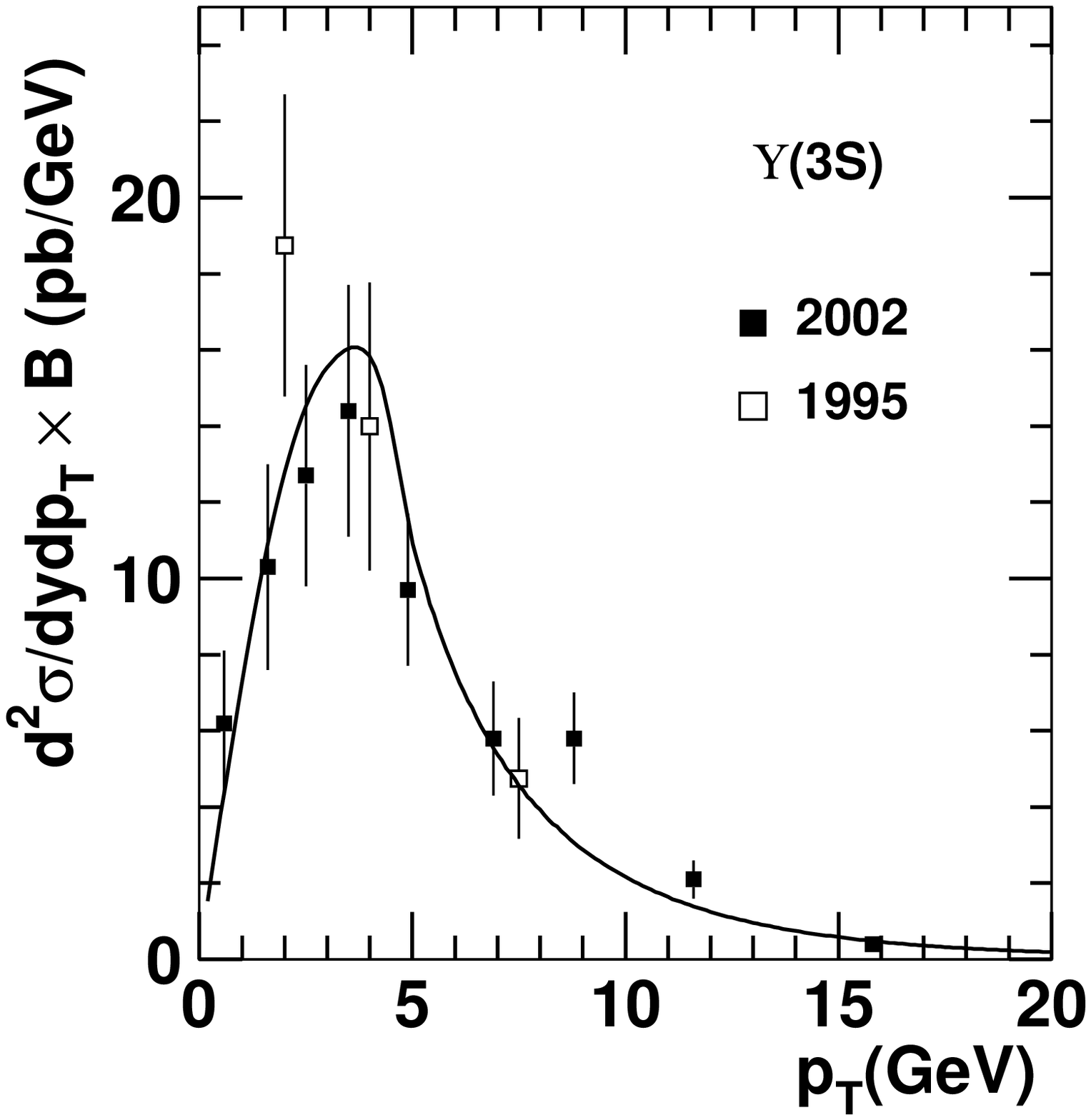, width=3.4cm}}
\caption{Differential cross sections times leptonic branching 
fractions $B$, at rapidity $y = 0$, as functions of transverse momentum 
for hadronic production of (a) $\Upsilon(1S)$, (b) $\Upsilon(2S)$, and 
(c) $\Upsilon(3S)$ at $\sqrt S = 1.8$~TeV along with published Run~1 data 
from the CDF collaboration.}
\label{fig6}
\end{figure}

The principal predictive power of our calculation is the shape of the 
$p_T$-distribution for the full $p_T$ region.  
In Fig.~\ref{fig6}, we present our calculation of the transverse  
momentum distribution for hadronic production of $\Upsilon(nS)$, $n =
1 - 3$.  The shapes of the $p_T$ distributions 
are consistent with experimental results. 
We find best fit values $K_r = 1.22 \pm 0.02$ and $p_{T_M} \sim 4.27$~GeV. 
The value of $C_{\Upsilon}$: 0.044, 0.040, and 0.041 for 
$\Upsilon(1S)$, $\Upsilon(2S)$, and $\Upsilon(3S)$, respectively,
is approximately independent of $M_{\Upsilon}$. 
\begin{figure}[ht]
\centerline{\psfig{file=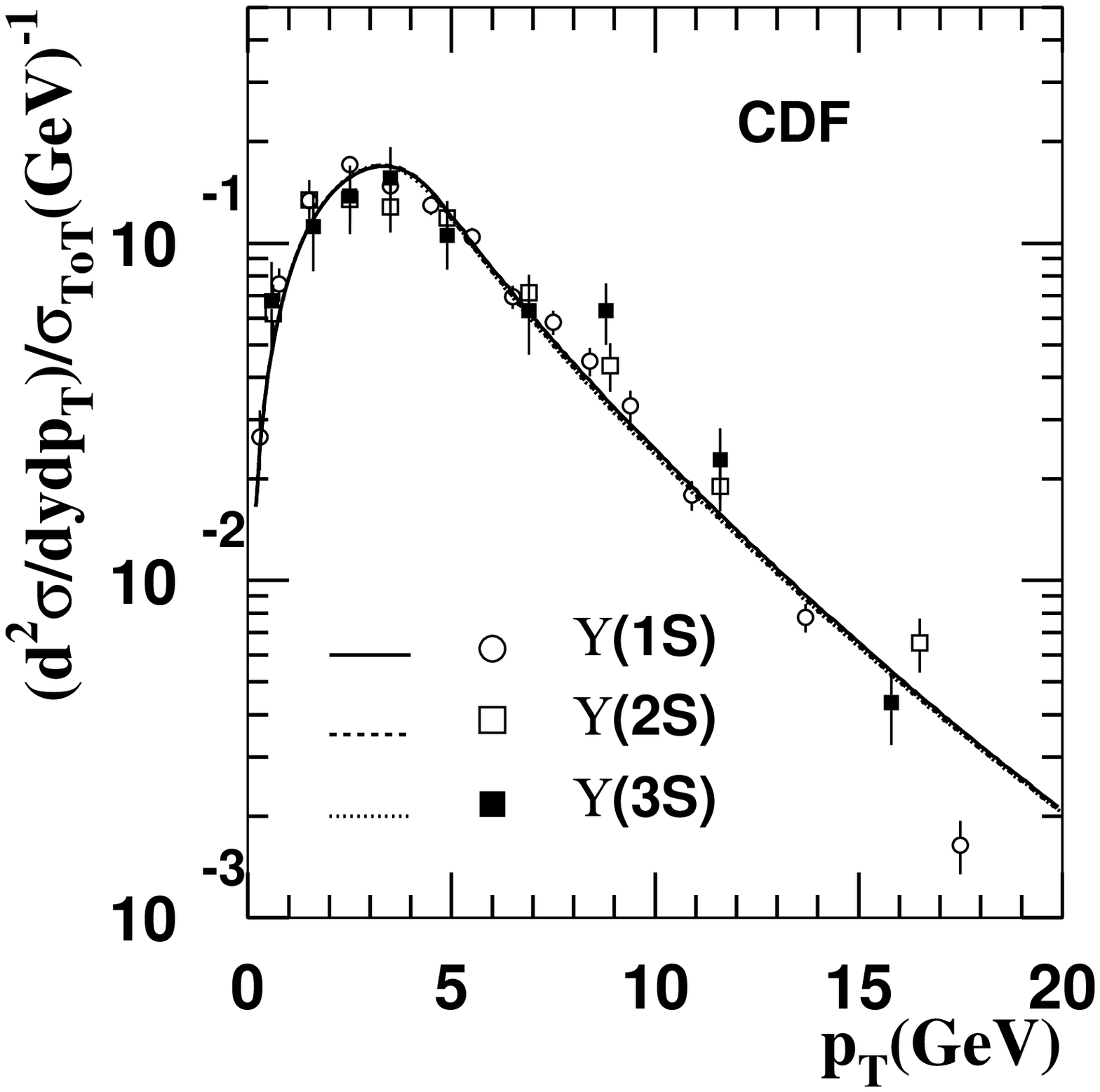,width=3.2cm} 
\hspace{0.2cm}\psfig{file=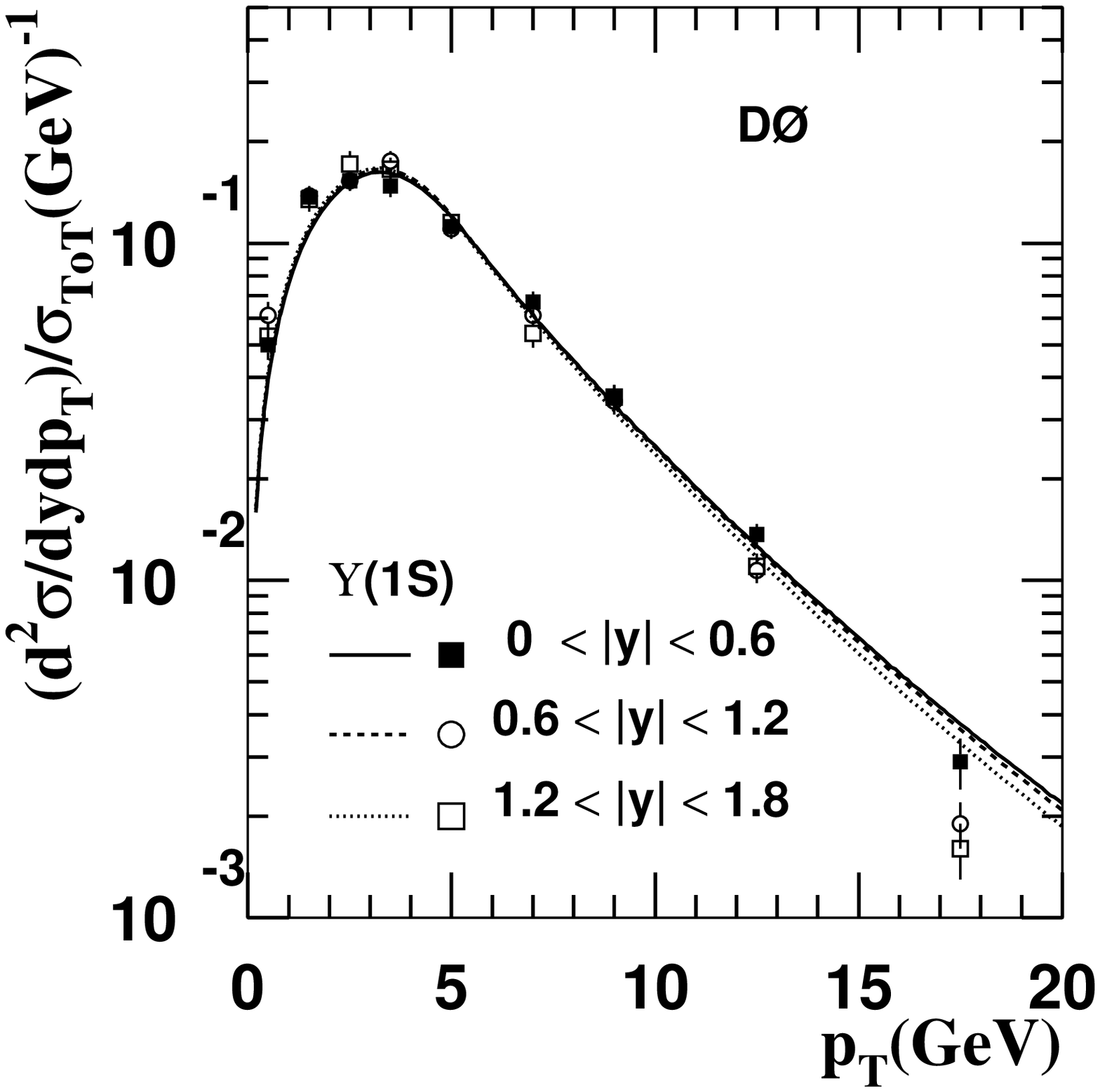,width=3.2cm}}
\caption{Normalized transverse momentum distribution for $\Upsilon$ production 
at a) $\sqrt S = 1.8$~TeV along with published Run~1 data
and b) $\sqrt S = 1.96$~TeV along with preliminary D\O\ Run~2 data.}
\label{fig7}
\end{figure}
The essential similarity of the production differential cross sections
for the three $\Upsilon(nS)$ states is illustrated in frame a) of Fig.~\ref{fig7}.  
The three theory curves are practically  indistinguishable. The curves in frame b) 
are our predictions for Run~2 at $\sqrt S = 1.96$~TeV for three different ranges 
of rapidity, along with preliminary D\O\ 
data~\cite{d0}.  Since the curves in Fig.~2 are normalized by the
integrated cross sections, dependence on the normalization parameters 
$C_{\Upsilon}$ cancels in the ratio. The good agreement with data is based on 
the choice of only two adjustable constants, the resummation enhancement factor 
$K_r$ and the matching point $p_{T_M}$. 
\section*{Acknowledgments} 
E.~L.~Berger is supported 
by the United States Department of Energy, Division of High Energy 
Physics, under Contract W-31-109-ENG-38.  J.~W.~Qiu is supported in part by 
the United States Department of Energy under Grant No. DE-FG02-87ER40371.  
Y.~Wang is supported in part by the United States Department of Energy
under Grant No. DE-FG02-01ER41155.


\begin{thebibliography}{0}
\bibitem{cdf}
CDF Collaboration, F. Abe {\em et al}, 
Phys.\ Rev.\ Lett.\  {\bf 75}, 4358 (1995); D.~Acosta {\it et al.} 
Phys.\ Rev.\ Lett.\  {\bf 88}, 161802 (2002).
\bibitem{Berger:2004cc}
E.~L.~Berger, J.~W.~Qiu, and Yili Wang,
arXiv:hep-ph/0404158.
\bibitem{d0}
D\O\ Collaboration, 
http://www-d0.fnal.gov/Run2Physics/WWW/results/b.htm, 
D\O\ Conference note 4523.
\bibitem{Collins:gx}
J.~C.~Collins, D.~E.~Soper, and G.~Sterman,
Adv.\ Ser.\ Direct.\ High Energy Phys.\  {\bf 5}, 1 (1988);
J.W.~Qiu and G.~Sterman, in preparation.
\bibitem{Amundson:1996qr}
J.~F.~Amundson {\em et al},
Phys.\ Lett.\ B {\bf 390}, 323 (1997)
[arXiv:hep-ph/9605295].
\bibitem{Collins:1984kg}
J.~C.~Collins, D.~E.~Soper, and G.~Sterman,
Nucl.\ Phys.\ B {\bf 250}, 199 (1985).

\end{thebibliography}
\end{document}